\documentclass[%
 reprint,
 amsmath,amssymb,
 aps,
]{revtex4-1}

\usepackage{graphicx}
\usepackage{dcolumn}
\usepackage{bm}
\usepackage{eqnarray,amsmath}

\begin{document}

\preprint{APS/123-QED}

\title{The Effects of Environment on Thermopower and Thermocurrent in Open Molecular Junctions}

\author{A. Eskandari-asl}
\email[Email: ]{a{\_}eskandariasl@sbu.ac.ir; amir.eskandari.asl@gmail.com}
\affiliation{Department of physics, Shahid Beheshti University, G. C. Evin, Tehran 1983963113, Iran}

\date{\today}

\begin{abstract}
In this work, we consider an open single level molecular junction with electrons coupled to a single frequency phonon mode, in which the phonons are also coupled to another heat bath. By applying a temperature gradient between the electrical leads and investigating thermocurrent in non-linear regime, it is shown that the environment can excite phonons which may result in either suppression or very significant enhancement of thermocurrent, depending on the onsite energy. Similarly, in the linear regime the effects of environment are seen in the Seebeck coefficient. 

\end{abstract}

\maketitle

\section{Introduction}\label{int}
In recent years, molecular junctions (MJ) which connect two large electronic reservoirs have been studied as building blocks for the future molecular electronics\cite{cuniberti,zebarjadi2012}. In addition to other special characteristics of MJs, their thermoelectric properties are of interest\cite{zimbovskaya2016}.

When there is a temperature difference between the two leads, an electrical current can pass through the MJ, so we have a thermopower. This is the so called Seebeck effect in MJs\cite{zimbovskaya2016,reddy2007}.
At the molecular size, in addition to the Coulomb interaction, the coupling of electrons to the vibrations of molecule is of a great importance. Indeed, for understanding thermoelectric properties, these effects have been considered extensively\cite{murphy2008,zimbovskaya2016,agarwal2014,monteros2014,chang2017,hartle2011,galperin2008inelastic,zotti2014heat,kruchinin2014,eskandari2018interplay}.

The MJ can be considered either thermally isolated from or connected to other heat baths\cite{de2016}.In theoretical studies, the latter is mostly modeled by considering the phonons on MJ to be totally thermalized. However, one can study a model in which a thermal bath is weakly coupled to MJ, which is different from both of these extreme cases\cite{eskandari2018influence}. It seems that in real experiments ideal isolation of a MJ is not a simple task and some parts of environment may still have weak thermal connection to the system. Especially, thermal connection to environment would be of great importance if we want to get closer to the practical applications of molecular electronics. Therefore, more theoretical studies should be done to understand the thermal effects of environment on MJs.

In this work, we study the effects of a thermal environment on the thermoelectric properties of a single level MJ which is coupled to single frequency phonons. We consider both non-linear and linear regimes.In the first case, the temperature gradient is noticeable compared to the lead temperatures and one should study thermocurrent\cite{leijnse2010,kim2016temperature,zimbovskaya2018}. On the other hand, for linear regime the temperature gradient is much smaller than the lead temperatures and the Seebeck coefficient is defined\cite{zimbovskaya2016,reddy2007,rincon2016}. We have shown that depending on the gate voltage, the environment can either suppress or very noticeably enhance thermocurrent. Actually, the environment excites phonons in MJ which in turn affects thermocurrent and Seebeck coefficient.

The paper is organized as follow. In Sec.\ref{mm}, we introduce the system Hamiltonian and the corresponding master equations(ME). Moreover, formulas for thermocurrent and Seebeck coefficient are also given. In Sec.\ref{nr}, we present our numerical results and discussions, and finally, Sec.\ref{con} concludes our work.

\section{Model and Method} \label{mm}
We consider a MJ which is connected to two spin-less electronic leads and a single frequency phonon mode\cite{kruchinin2014,eskandari2018influence}. This phonon mode is also coupled to another thermal phononic bath. The Hamiltonian of this system is 
\begin{eqnarray}
&&\hat{H}=\hat{H}_{m}+\hat{H}_{leads}+\hat{H}_{tun}+\hat{H}_{bath}+\hat{H}_{m-bath},
\label{hamt}
\end{eqnarray}
\begin{eqnarray}
&&\hat{H}_{m}=\epsilon_{0} \hat{n}_{d}+\Omega \hat{b}^{\dag}\hat{b}+\lambda \Omega \hat{n}_{d} \left(\hat{b}+ \hat{b}^{\dag} \right),
\label{hm}
\end{eqnarray}
\begin{eqnarray}
&&\hat{H}_{leads}=\sum_{k,\alpha \in \left\lbrace R,L\right\rbrace } \epsilon_{k,\alpha} \hat{a}^{\dag}_{k\alpha} \hat{a}_{k\alpha} ,
\label{hleads}
\end{eqnarray}
\begin{eqnarray}
&&\hat{H}_{tun}=\sum_{k,\alpha \in \left\lbrace R,L\right\rbrace } V_{k\alpha} \hat{c}^{\dag} \hat{a}_{k\alpha}+ h.c. ,
\label{htun}
\end{eqnarray}
\begin{eqnarray}
&&\hat{H}_{bath}=\sum_{\nu} \Omega_{\nu} \hat{b}^{\dag}_{\nu} \hat{b}_{\nu},
\label{hbath}
\end{eqnarray}  
and
\begin{eqnarray}
&&\hat{H}_{m-bath}=\sum_{\nu} \gamma_{\nu} (\hat{b}^{\dag}+\hat{b})(\hat{b}^{\dag}_{\nu}+\hat{b}_{\nu}),
\label{hm-b}
\end{eqnarray}  

where $ \hat{c} $ ($ \hat{c}^{\dag} $) is the annihilation (creation) operator of electrons on MJ, $ \hat{n}_{d}=\hat{c}^{\dag}\hat{c} $ is the number operator and $ \epsilon_{0} $ is the onsite energy of electrons on the MJ. $ \hat{b} $($ \hat{b}^{\dag} $) is the annihilation (creation) operator of phonons on MJ, $ \Omega $ is the phonon frequency and $ \lambda $ determines electron-phonon coupling. Moreover, $ \hat{a}_{k\alpha} $ ($ \hat{a}^{\dag}_{k\alpha} $) annihilates (creates) an electron in the state $ k $ of the lead $ \alpha $ ($ \alpha=R,L $), and  $ V_{k,\alpha} $ determines the electron hopping between MJ and the leads. $ \hat{b}_{\nu} $ ($ \hat{b}^{\dag}_{\nu} $) is the annihilation (creation) operator of mode $ \nu $ of the thermal bath, $ \Omega_{\nu} $ is the energy of this mode, and $ \gamma_{\nu} $ determines the coupling strength of this mode with MJ phonons.

Similar to a former work\cite{eskandari2018influence}, performing the Lang-Firsov transformation as $ e^{\hat{S}} \hat{H} e^{-\hat{S}} $ (where $ \hat{S}\equiv \lambda \hat{n}_{d} \left( \hat{b}^{\dag}-\hat{b} \right) $ ), renormalizing the onsite energy to $ \epsilon=\epsilon_{0} - \lambda^{2} \Omega $ and following the standard steps for deriving a Markovian ME in the limit of weak lead to MJ coupling, one can obtain the dynamics of the density matrix (DM) of the system (where by system we mean the electrons and phonons of MJ). After doing straight forward calculations, the rate of change of diagonal elements of the electron-phonon DM is obtained as
\begin{eqnarray}
&&\frac{d}{dt} P_{0m}=\sum_{m^{\prime},\alpha } \Gamma_{\alpha} \left( \left[ 1-f_{\alpha}\left( \Omega \left( m^{\prime}-m\right)  +\epsilon \right)\right] \vert \hat{X}_{mm^{\prime}} \vert^{2} P_{1,m^{\prime}}\right.\quad\nonumber\\
&&\left.-f_{\alpha}\left( \Omega \left( m^{\prime}-m\right)  +\epsilon \right) \vert \hat{X}_{mm^{\prime}} \vert^{2} P_{0m} \right)+\hat{\mathcal{L}}_{b} (P_{0m}),
\label{dp0}
\end{eqnarray}  

\begin{eqnarray}
&&\frac{d}{dt} P_{1m}=\sum_{m^{\prime},\alpha } \Gamma_{\alpha} \left( f_{\alpha}\left( \Omega \left(m- m^{\prime}\right)  +\epsilon \right) \vert \hat{X}_{m^{\prime}m} \vert^{2} P_{0m^{\prime}}\right.\nonumber\\
&&\left.-\left[ 1-f_{\alpha}\left( \Omega \left(m- m^{\prime}\right)  +\epsilon \right)\right] \vert \hat{X}_{m^{\prime}m} \vert^{2} P_{1,m}\right)+\hat{\mathcal{L}}_{b} (P_{1m}), \qquad
\label{dp1}
\end{eqnarray}  
where $ P_{im} $ ($ i=0,1 $) represent diagonal elements of DM and determine the probability of having $ i $ electrons and $ m $ phonons in MJ. Moreover, $ \hat{X}\equiv\exp [\lambda(\hat{b}-\hat{b}^{\dag})] $, and $ f_{\alpha}(\omega)=\frac{1}{e^{\beta_{\alpha}(\omega-\mu_{\alpha})}+1} $ is the Fermi distribution of lead $ \alpha $, in which $ \mu_{\alpha} $ is chemical potential of the lead and $ \beta_{\alpha} $ is its inverse temperature. $ \Gamma_{\alpha} $ determines the tunneling rate of electrons between MJ and lead $ \alpha $, which is defined to be $ \Gamma_{\alpha}(\omega)=\sum_{k} 2 \pi \vert V_{k\alpha} \vert^{2} \delta(\epsilon_{k\alpha}-\omega) $. In wide band limit(WBL), we take $ \Gamma_{\alpha} $ to be independent of $ \omega $. $ \hat{\mathcal{L}}_{b} (P_{im}) $ is
\begin{eqnarray}
\hat{\mathcal{L}}_{b} (P_{im})&=& \Gamma_{p} \left[ 1+N_{bath}\left( \Omega \right) \right] \left[\left( m+1\right)  P_{i,m+1}-m P_{im}\right] +\nonumber \\
&&\Gamma_{p} N_{bath}\left( \Omega \right) \left[m P_{i,m-1}-\left( m+1\right)  P_{i,m}\right],
\label{lb}
\end{eqnarray}
in which $ N_{bath}\left( \Omega \right)=\frac{1}{e^{\beta_{b}\Omega}-1} $ is the number of phonons with frequency $ \Omega $ in the thermal bath (given by Bose-Einstein distribution function), in which $ \beta_{b}=\frac{1}{k_{B} T_{b}} $ is the inverse temperature of the phonon bath. Moreover, $ \Gamma_{p}=\sum_{\nu} 2 \pi \gamma_{\nu}^{2} \delta(\Omega-\Omega_{\nu}) $, determines the thermalization rate of MJ phonons.  

The number of electrons in MJ is $ N_{e}=\sum_{m} P_{1m} $. Moreover, the particle current from lead $ \alpha $ to the MJ is obtained as \cite{eskandari2018influence}
\begin{eqnarray}
I_{\alpha}&=&\Gamma_{\alpha}\sum_{mm^{\prime}} \left(  -\left[ 1-f_{\alpha}\left( \Omega \left( m^{\prime}-m\right)  +\epsilon \right)\right] \vert \hat{X}_{mm^{\prime}} \vert^{2} P_{1,m^{\prime}}+ \right.\nonumber\\
&&\left.f_{\alpha}\left( \Omega \left(m- m^{\prime}\right)  +\epsilon \right) \vert \hat{X}_{m^{\prime}m} \vert^{2} P_{0m^{\prime}}\right) ,
\end{eqnarray}
and the total current passing through the MJ is $ I=(I_{L}-I_{R})/2 $. By considering $ \Gamma_{L}=\Gamma_{R}=\Gamma $, the electrical current through MJ in steady state is given by
\begin{eqnarray}
I&=&\frac{1}{2}\Gamma \sum_{mm^{\prime}}   \vert \hat{X}_{m^{\prime}m} \vert^{2} \left(P_{0m^{\prime}}+P_{1m} \right)\times \nonumber\\
&& \left[f_{L}\left( \Omega \left(m- m^{\prime}\right) +\epsilon \right)-f_{R}\left( \Omega \left(m- m^{\prime}\right) +\epsilon \right)\right] ,\quad.
\label{ecurr}
\end{eqnarray}

One interesting quantity in our work is the thermopower which is determined by the Seebeck coefficient. In order to compute Seebeck coefficient we consider the equilibrium case where we have no temperature gradient and the chemical potential of the leads are equal to each other. In this case, we can use the notation $ f_{L}=f_{R}=f $ and $  \mu_{L}=\mu_{R}=\mu $. If negligible temperature gradient, $ \delta T $, and bias voltage, $ \delta V $, are applied, to linear order the current through MJ becomes
\begin{eqnarray}
&&\delta I=\frac{1}{2}\Gamma \sum_{mm^{\prime}}   \vert \hat{X}_{m^{\prime}m} \vert^{2} \left(P_{0m^{\prime}}+P_{1m} \right)\times \nonumber\\
&& \left[\frac{\partial f\left( \Omega \left(m- m^{\prime}\right) +\epsilon \right)}{\partial \mu} \delta V+\frac{\partial f\left( \Omega \left(m- m^{\prime}\right) +\epsilon \right)}{\partial T} \delta T\right] ,\qquad
\label{di}
\end{eqnarray} 
The Seebeck coefficient is defined as $ S=-\delta V / \delta T  $, in the limit of vanishing $ \delta T $, provided that the current vanishes, i.e., $ \delta I=0 $. Using Eq.\ref{di}, this results in 
\begin{eqnarray}
S=\frac{\sum_{mm^{\prime}} \vert \hat{X}_{m^{\prime}m} \vert^{2} \Pi_{m^{\prime}m} E_{m^{\prime}m} g\left(  E_{m^{\prime}m} \right)  }{T_{l} \sum_{mm^{\prime}} \vert \hat{X}_{m^{\prime}m} \vert^{2} \Pi_{m^{\prime}m} g\left(  E_{m^{\prime}m} \right)}
\label{seebeck}
\end{eqnarray}
where $ T_{l} $ is the common temperature of the leads ($ \beta=1/k_{B}T_{l} $), $ \Pi_{m^{\prime}m}=\left(P_{0m^{\prime}}+P_{1m} \right) $, $E_{m^{\prime}m}= \left( \Omega \left(m- m^{\prime}\right) +\epsilon-\mu  \right) $, and $ g(\omega)=e^{\beta\omega}/\left( e^{\beta\omega}+1\right) ^{2} $.

In some special cases we can approximate the value of $ S $. For the case where $ \epsilon \ll \mu $ ($ \epsilon \gg \mu $) and the temperatures of leads and bath are low enough, the MJ is filled with one electron (empty of electrons) and approximately there are no phonons. As a result, all $ P_{im} $s are zero except $ P_{1,0} $ ($ P_{0,0} $) which is equal to 1. Inserting this in Eq.\ref{seebeck}, results in $ S=(\epsilon-\mu) /T_{l} $. Additionally, if the electrons weren't coupled to phonons at all (i.e., $ \lambda=0 $), the only non-zero terms would be those with $ m=0 $, and from Eq.\ref{seebeck},  the Seebeck coefficient is again $ S=(\epsilon-\mu) /T_{l} $. Moreover, for the case where $ \epsilon\approx \mu $ and the temperatures are low enough, $ g\left( \Omega \left(m- m^{\prime}\right) +\epsilon-\mu\right) $ is very small for $ m\neq m^{\prime} $. Using Eq.\ref{seebeck}, one still obtains $ S=(\epsilon-\mu) /T_{l} $. It is noticeable that these approximations can be used as a check of the correctness of some of the numerical results.

\section{Numerical Results} \label{nr}
In this section we represent our numerical results. We work in a system of units in which $ e=\hbar=1 $. Also, the Boltzmann constant, $ k_{B} $, is taken to be 1. We set the phonon frequency to be our energy unit, i.e., $ \Omega=1 $. These automatically set our units of temperature, electrical current, and Seebeck coefficient, whose units are $ \frac{\hbar \Omega}{k_{B}} $, $ e \Omega $ and $ \frac{k_{B}}{e} $, respectively. Moreover, $ \lambda=1 $ and no bias voltage is applied between the leads, so that $ \mu_{L}=-\mu_{R}=0 $. Finally, the tunneling rates between MJ and the leads are assumed to be $ \Gamma_{L}=\Gamma_{R}=0.1 \Omega=0.1 $.

\begin{figure}  
\includegraphics{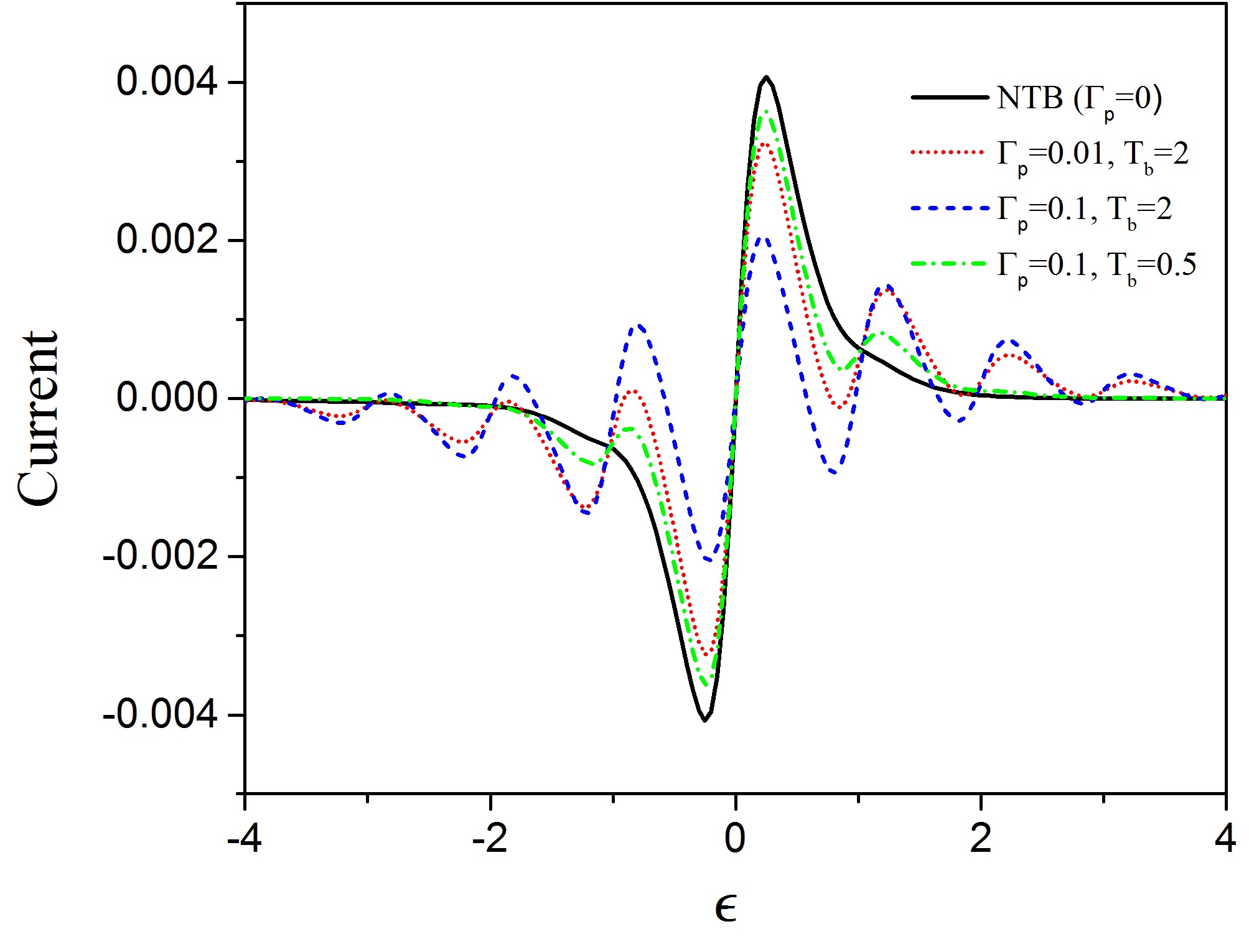}
\caption{\label{fig1} The current as a function of onsite energy of MJ, for the NTB case (solid-black curve) and several cases of having a thermal bath coupled to the MJ. It is seen as the coupling strength($\Gamma_p$) and/or the temperature of thermal bath($T_{b}$) is increased, the fluctuations in thermocurrent which are the finger prints of phonon side-bands, get more significant. The temperature of the right(left) lead is $ T_{R}=0.1 $ ($ T_{L}=0.3 $) and the temperatures of thermal bath and the coupling strengths are indicated in the plot.Units are discussed in the text.}
\end{figure}

First, we study the environmental effects on thermocurrent. For this purpose, we need to apply a temperature difference between the leads. We consider the right(left) lead to be at the temperature $ T_{R}=0.1 $ ($ T_{L}=0.3 $). In Fig.\ref{fig1} the thermocurrent is depicted as a function of onsite energy,$ \epsilon  $, for several thermal baths. For the case where we have no thermal bath coupled to the MJ (the NTB case ,solid-black curve), it is seen that the current changes sign by $ \epsilon $, which stems from changing the current carriers from holes to electrons. This phenomenon is well known in single level quantum bridges\cite{zimbovskaya2016}. Since the temperatures of leads are low and the electrical current is small, the phonons can not be excited in the MJ and their sidebands have negligible effects on thermocurrent. It is noteworthy that for typical phonon energy values of the order of  10 meV, the temperature difference here will become of the order of 10 K and the maximum current is also of the order of 10 nA. 

Other curves in Fig.\ref{fig1} correspond to the cases where we have coupling between the MJ and thermal bath. As it is seen, by increasing the coupling strength and/or the bath temperature, the thermocurrent gets more alternating and the distance between the current peaks is almost equal to the phonon frequency. The thermal bath excites phonons in the MJ which results in increasing the effect of phonon sidebands. It should be noted that coupling to thermal bath decreases the maximum value of thermocurrent, which can be understood by noting that the excited phonons block the electrons and reduce the current, similar to the well-known phenomenon of Franck-Condon blockade. On the other hand, for the cases where $ \epsilon $ is in the proximity of $ m \Omega $ ($ m\neq 0) $, the value of thermocurrent can be noticeably increased by the environment, since the excited phonons open new transport channels for electrons. That is, electrons can combine with phonons to create polarons that can pass trough the junction.

\begin{figure}  
\includegraphics{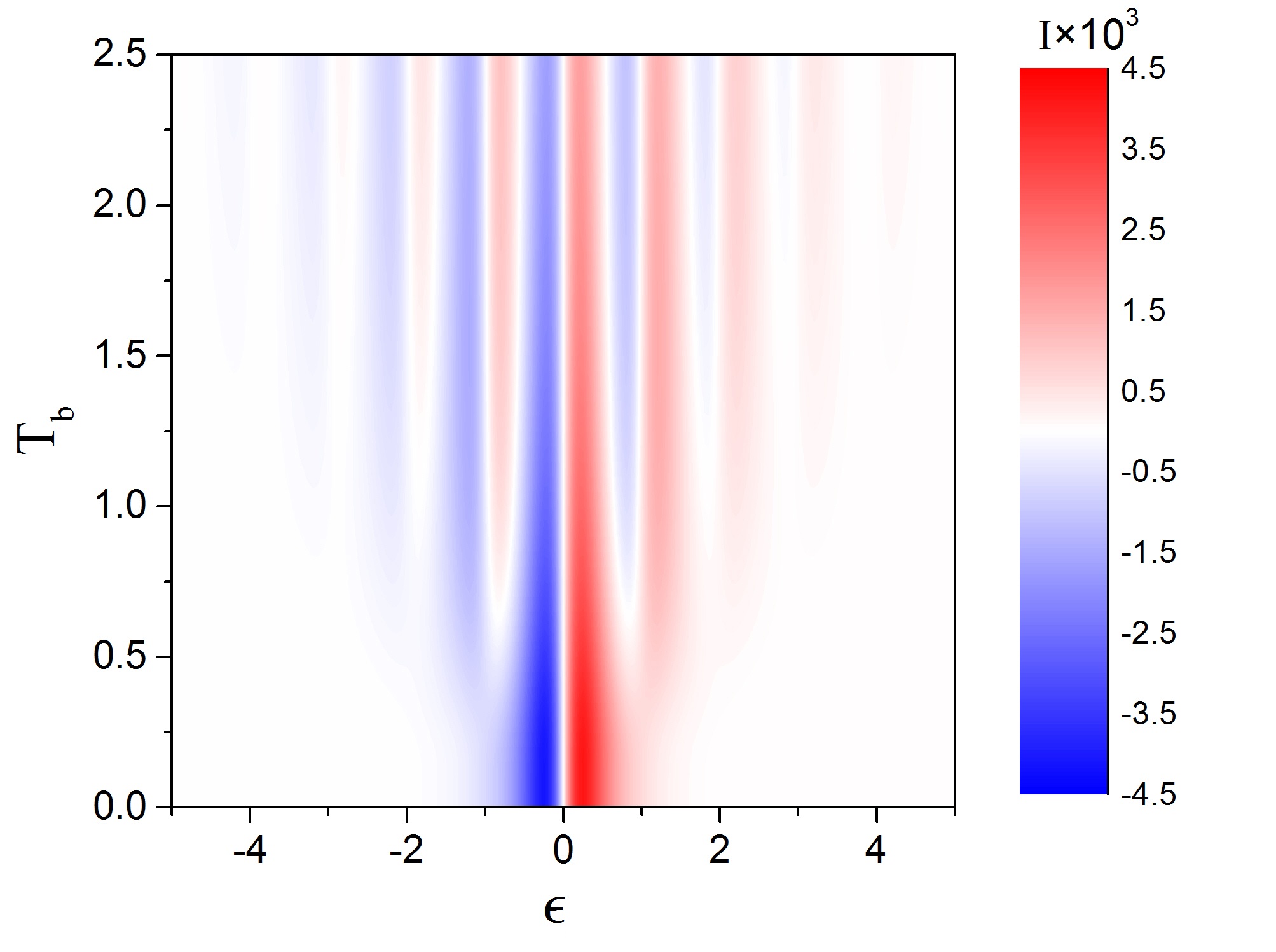}
\caption{\label{fig2} The current as a function of onsite energy of MJ and the bath temperature. The temperature of the right(left) lead is $ T_{R}=0.1 $ ($ T_{L}=0.3 $). By increasing the bath temperature, more side-bands get involved and we have more fluctuations in the thermocurrent.Units are discussed in the text.}
\end{figure}

In Fig.\ref{fig2}, in a two dimensional plot we show the behavior of thermocurrent as a function of onsite energy and bath temperature, with the same temperature gradient as Fig.\ref{fig1}. As it is seen, by increasing the bath temperature,$ T_{b} $, more peaks and valleys emerge in thermocurrent, while the maximum current decreases. 

Next, we investigate the behavior of the thermopower, or the Seebeck coefficient, for the case where the leads temperatures are equal (the common temperature is $ T_{l} $) and as before, we have no voltage bias . However, here some technical issues emerge which have to be resolved.  

The first problem arises for the NTB case, for which it turns out that the steady state solution of Eqs. \ref{dp0} and \ref{dp1} is not unique. In order to clarify the situation, consider we have $ \epsilon \ll 0 $ and $ T_{l} $ is not very high. If we choose an empty junction as our initial state, one electron rapidly jumps into the junction and this would excite phonons. As a result, the effective temperature of the junction (which is determined by the phonon population \cite{eskandari2018influence}) can be much higher than that of the leads. On the other hand, if we start with a junction that is occupied with one electron, no more hopping can take place and the expected value of phonon number is very close to zero.  Similar reasoning can be applied for the case where $ \epsilon \gg 0 $. Note that the electrical current in these situations is almost zero, so that our former result for NTB case is valid. In order to avoid this apparent bi-stability, we consider a thermal bath with the same temperature as the leads that is coupled to our junction with a small coupling strength (0.01 in our numerical calculations).

The second problem is for small lead temperatures and $ \mid \epsilon / T_{l} \mid \gg 0 $ , where for some $ m $ and $ m^{\prime} $, $ g\left(  E_{m^{\prime}m} \right)$ can be orders of magnitude larger than  $ g\left( \epsilon\right) $ (see Eq.\ref{seebeck}). Consequently, a very small numerical error in vanishing $ P_{im} $s can result in a completely incorrect value for $ S $. This problem can be resolved by choosing the initial conditions in such a way that the zero $ P_{im} $s don't grow at all. This can be achieved by considering an initially full (empty) junction for the case where $ \epsilon <0 $ ($ \epsilon >0 $). 

\begin{figure}  
\includegraphics{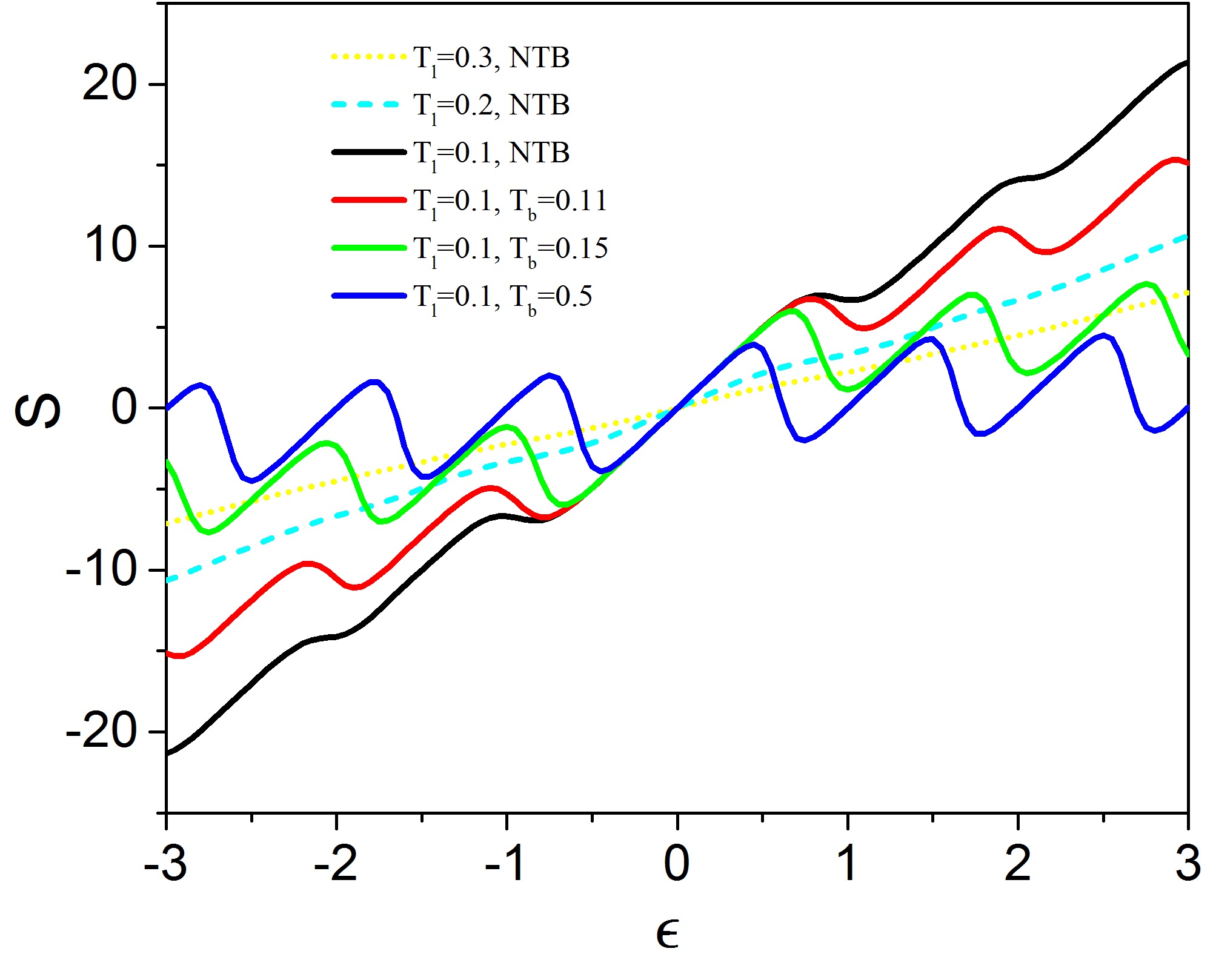}
\caption{\label{fig3} The Seebeck coefficient, $ S $, as a function of onsite energy of MJ, for different lead and bath temperatures ($ T_{l} $ is the common temperature of leads  and $ T_{b} $ the bath temperature). It is seen that a hot bath can excite the phonons and open the transport channels corresponding to phonon side-bands, so change the behavior of Seebeck coefficient drastically.Units are discussed in the text.}
\end{figure}  

In Fig.\ref{fig3}, we depict $ S $ as a function of $ \epsilon $ for several situations. For the NTB case, the behavior is almost linear (which is consistent with our approximations after Eq.\ref{seebeck}), with small deviations caused by the e-ph coupling. However, as the $ T_{b} $ gets higher, the effects of the opened phononic channels become more important.The linear behavior of $ S $ is not restricted to the vicinity of $ \epsilon=0 $ anymore, but around any open transport channel we can have a linear relationship. By increasing the bath temperature,$ T_{b} $, phonon channels at the energies of $ \epsilon +m \Omega$ ($ m $ is an integer) become more transparent and the linear behavior gets stronger, consequently, the plot of $ S $ vs $ \epsilon $ approaches a zigzag form. One other feature of these curves is that they are all odd with respect to $ \epsilon $. It is physically understood by noting that the roles of electrons and holes interchange by changing the sign of $ \epsilon $, which would result in changing the sign of thermopower.As it is noted at the beginning of this section, the unit of $ S $ in our calculations is $\frac{k_{B}}{e}=8.62\times 10^{-5} V/K $, so that our typical maximum values are of the order of $ 10^{-4}\sim 10^{-3} $ in SI units.

\section{Conclusions} \label{con}
In conclusion, we considered thermoelectric properties of a single level MJ with single frequency phonons that is also coupled to a thermal environment. At first, by applying a temperature difference between the leads we obtained the thermocurrent as a function of onsite energy of MJ(a gate voltage) for zero bias voltage. It is shown that the coupling to the environment can excite phonons in the MJ. As a result, the main peaks that would exist in the absent of coupling to environment(the NTB case) are suppressed, which shows the phonons block electrons (similar to Franck-Condon blockade). On the other hand, excited phonons can open new side-bands as transport channels, so we have a set of new peaks and valleys in thermocurrent. At such onsite energies, the thermocurrent is very noticeably enhanced by environment.

We also computed the Seebeck coefficient as a function of onsite energy. For the NTB case, it behaves almost linearly with a slope that is inversely proportional to the common temperature of the leads. By increasing the coupling to the heat bath, again the peaks and valleys corresponding to the opened side-bands appear.    
  
 


\bibliography{apssamp}

\end{document}